\documentclass[amsmath,amssymb,pra,twocolumn]{revtex4-1}

\usepackage{graphicx}
\usepackage{dcolumn}
\usepackage{bm}
\usepackage{hyperref}

\begin{document}

\title{Lamb-peak spectrum of the HD (2-0) P(1) line}

\author{M. L. Diouf, F. M. J. Cozijn, K.-F. Lai, E. J. Salumbides, W. Ubachs}
 \email{w.m.g.ubachs@vu.nl}
\affiliation{%
Department of Physics and Astronomy, LaserLab, Vrije Universiteit\\
 De Boelelaan 1081, 1081 HV Amsterdam, The Netherlands 
}%

\date{\today}

\begin{abstract}
A saturation spectroscopy measurement of the P(1) line of the ($2-0$) band in HD is performed in a sensitive cavity-enhanced optical setup involving frequency comb calibration. The spectral signature is that of a Lamb-peak, in agreement with a density-matrix model description involving 9 hyperfine components and 16 crossover resonances of $\Lambda$-type. Comparison of the experimental spectra with the simulations yields a rovibrational transition frequency  at 209,784,242,007 (20) kHz. Agreement is found with a first principles calculation in the framework of non-adiabatic quantum electrodynamics within 2$\sigma$, where the combined uncertainty is fully determined by theory.

\end{abstract}
\maketitle

\section{introduction}

In the recent decade the hydrogen molecule has become a benchmark system for testing quantum electrodynamics (QED) and probing physics beyond the Standard Model~\cite{Ubachs2016}. Such tests can be accomplished by comparison between accurate measurements, of e.g. dissociation energies of the H$_2$ molecule~\cite{Liu2009,Cheng2018,Holsch2019}, with non-adiabatic calculations of the hydrogen molecule based on a 4-particle variational framework and including relativistic and QED-terms up to $m\alpha^6$~\cite{Puchalski2019b}. Alternatively, tests were performed on splittings between rovibrational levels in the H$_2$ and D$_2$ molecules, measured at increasing accuracy~\cite{Cheng2012,Tan2014,Mondelain2016}. In view of the very small oscillator strengths of the quadrupole transitions probed in these homonuclear species, measurements were performed from Doppler-broadened and collision-broadened spectra limiting the accuracy. The application of sophisticated line-shape models in combination with high signal-to-noise ratio has nevertheless pushed the accuracy to the sub-MHz regime~\cite{Wcislo2018,Zaborowski2020}.

In case of the heteronuclear HD isotopologue the molecule exhibits a small electric dipole moment~\cite{Trefler1968,Pachucki2008} giving rise to an electric dipole absorption spectrum which was discovered by Herzberg~\cite{Herzberg1950} and investigated over the years under conditions of Doppler broadening, for the vibrational bands~\cite{Durie1960,Lin2000,Kassi2011,Fasci2018}, as well as for pure rotational transitions~\cite{Evenson1988,Ulivi1991}.

Recently, intracavity absorption techniques were explored in which saturation of the R(1) transition in the ($2-0$) overtone band was demonstrated~\cite{Tao2018,Cozijn2018}, delivering a reduction in linewidth of over a thousand compared to the Doppler-broadened lines. While these studies produced highly accurate transition frequencies for the R(1) ($2-0$) line in HD, a large discrepancy was found between the two studies. A reanalysis at improved signal-to-noise ratio obtained with the NICE-OHMS (Noise-Immune Cavity-Enhanced Optical-Heterodyne Molecular Spectroscopy) technique by the Amsterdam team showed that the line shape appeared as dispersive-like~\cite{Diouf2019}. This phenomenon was explained as a result of underlying hyperfine structure involving a large number of crossover resonances in the saturation spectrum. A simulation of the spectrum was produced via a density-matrix model with optical Bloch equations that well reproduced the dispersive line shape.
An extended analysis by the Hefei group, using three different cavity-enhanced techniques, resulted in a similar dispersive line shape~\cite{Hu2019}. That was however interpreted as a Fano line shape, caused by an interference between the rovibrational R(1) transition and an underlying continuum, reminiscent of the Fano line shape observed by the interference between transitions in the fundamental vibration of HD and broad collisionally-induced continuum resonances~\cite{McKellar1973}. These discrepancies on transition frequencies and diverging explanations for the observed line shapes call for extended measurements, in particular since rovibrational transitions in the heteronuclear HD molecule, allow for extreme precision and constitute an excellent test ground for molecular QED.

\section{experimental}

The P(1) line at $1.43\,\mu$m was measured  under conditions of saturation at a number of pressures in the range 0.5 Pa to 16 Pa, using the NICE-OHMS setup, details of which were previously documented~\cite{Cozijn2018,Diouf2019}. Some notable changes were made for the measurement of P(1). Firstly, the modulation frequency has been doubled to twice the free-spectral-range (FSR) of the cavity at 2 times 305 MHz. This was necessary to avoid absorption from a neighbouring water transition giving rise to a Lamb-dip from the carrier-sideband saturation occurring at half the FSR detuning.
Secondly, a dominant source of residual amplitude modulation due to spurious background etaloning was identified and eliminated, which allowed multiple scans to be much more effectively averaged. This averaging for attaining a good signal-to-noise ratio was made possible through the lock of the spectroscopy laser to a frequency comb; averages over up to 60 recordings were taken.
Thirdly, a liquid nitrogen cryotrap was installed to continuously pump outgassing water vapor during the measurements, which otherwise gave a significant background drift.

\section{Hyperfine structure}

\begin{figure}
\begin{center}
\includegraphics[width=0.8\linewidth]{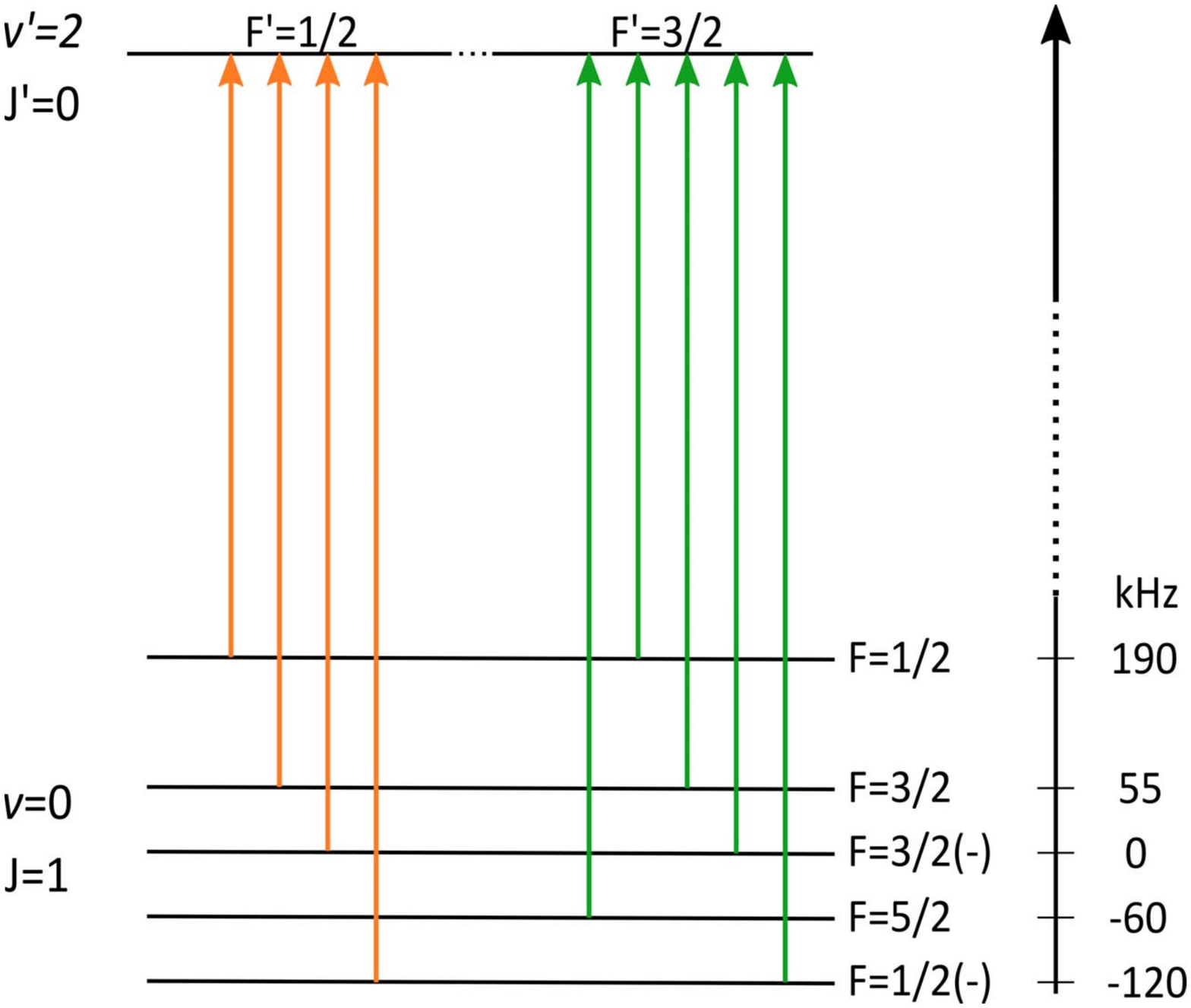}
\caption{\label{Hyperfine}
The hyperfine level structure with the five sublevels of the $v=0, J=1$ ground state, plotted on a scale corresponding to measurement~\cite{Ramsey1957}, and the two degenerate levels in the $v=2, J=0$ state}
\end{center}
\end{figure}

For an interpretation of the measured spectra of the P(1) line its hyperfine structure  must be considered, which is much simpler than that of the R(1) line measured previously~\cite{Diouf2019}. The  $J=1$ ground level is split into 5 components via the coupling of the $J=1$ rotational angular momentum to the nuclear spins $I_P=1/2$ for the proton and $I_D=1$ for the deuteron, and yields, in energetic order from high to low, $F=1/2(+)$, $F=3/2(+)$, $F=3/2(-)$, $F=5/2$, and $F=1/2(-)$ sublevels. Here the (+) and (-) refer to the highest and lowest levels with the same $F$-quantum number. The hyperfine splitting in $J=1$ was accurately measured by Ramsey and coworkers~\cite{Ramsey1957,Quinn1958}. The $v=2, J=0$ excited state is split into two components $F=1/2$ and $F=3/2$ that are however degenerate within a kHz; we assume the splitting to be approximately similar to that of $J=0$ in the $v=0$ ground state, where it was calculated as 45 Hz~\cite{Puchalski2018b}. The hyperfine level structure of the P(1) is plotted in Fig.~\ref{Hyperfine}. Our analysis of the hyperfine structure of this P(1) line is found to be in excellent agreement with that of other recent analyses~\cite{Puchalski2020,Dupre2020}.

\begin{figure}
\begin{center}
\includegraphics[width=0.9\linewidth]{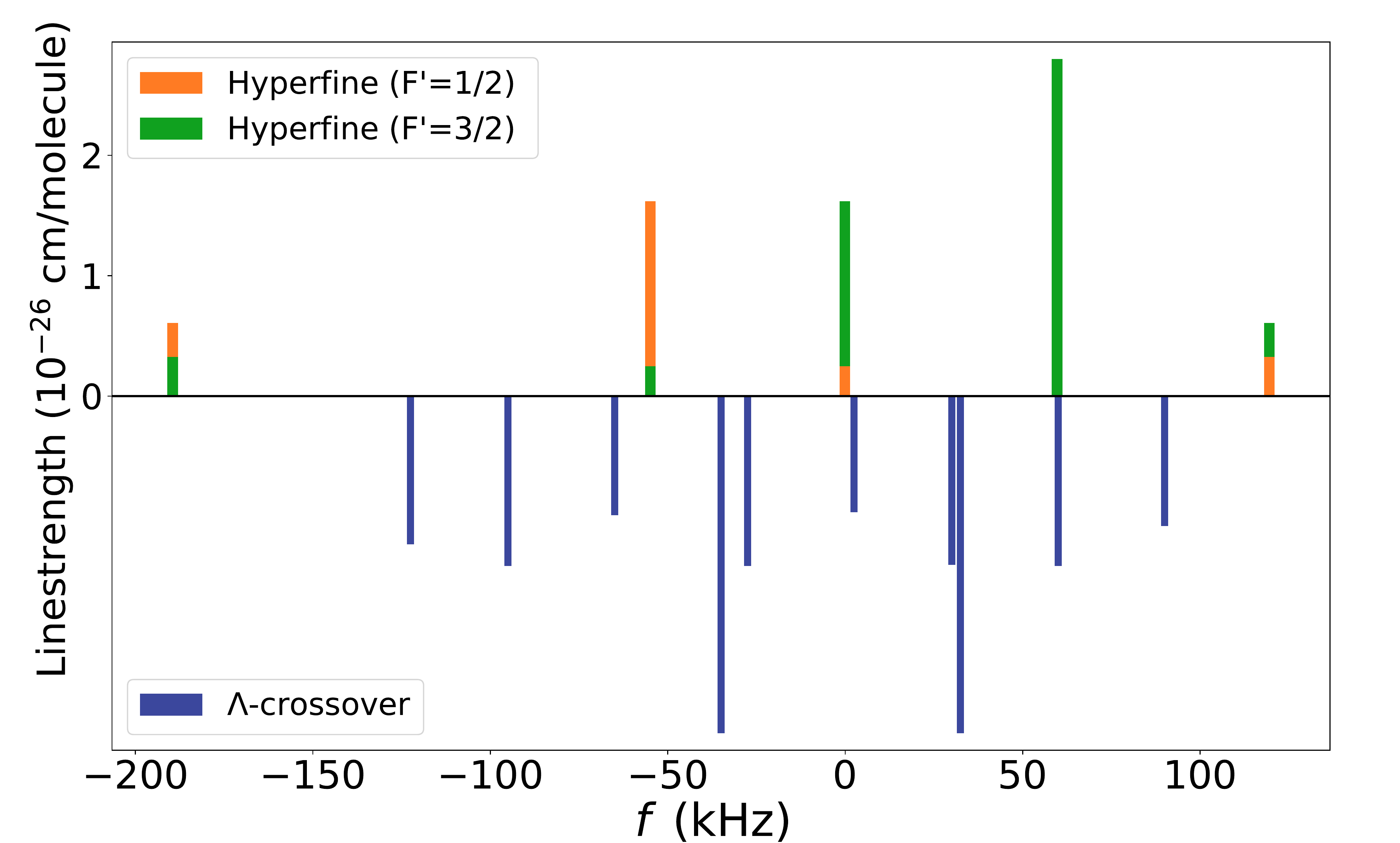}
\caption{\label{P1_stick}
Stick spectrum displaying hyperfine sub-components (of the 9 only 5 are non-degenerate) and crossover resonances (of the 16 only 10 are non-degenerate) of the P(1) line of HD on a frequency scale, where "0" is the pure rovibrational transition frequency.}
\end{center}
\end{figure}

The hyperfine structure results in 9 possible transitions connecting ground and excited state. Combination of these ground and excited states gives rise to 16 possible crossovers in the saturation spectrum, which are all of $\Lambda$-type due to near-degeneracy of the excited state. These resonances are plotted as a stick spectrum in Fig.~\ref{P1_stick}; note that in view of the degeneracy of the $F=1/2$ and $F=3/2$ upper levels not all individual components are visible. The relative intensities of the direct hyperfine components are calculated via angular momentum algebra~\cite{Pgopher}, where the absolute scale is matched to the measured value of the line intensity of the Doppler-broadened line~\cite{Kassi2011}. Intensities of crossover resonances are estimated via $\sqrt{I_iI_j}$ with $I_i$ and $I_j$ the intensities of the crossing hyperfine resonances. Note that these intensities are not used in the density-matrix model discussed below.
The hyperfine structure of P(1), as displayed in Fig.~\ref{P1_stick}, shows the sharp contrast to the R(1) line where both V-type and $\Lambda$-type crossover resonances contribute to the spectrum.

Note that V-type crossovers show as a Lamb-dip, as a common ground state couples to two excited states causing depletion of population, and therewith reduced absorption. However, $\Lambda$-type crossover resonances couple two ground states to a common excited state. This may lead to increased population in one ground state, resulting in increased absorption, and hence a Lamb-peak. In the R(1) spectrum of HD, where both V-type and $\Lambda$-type crossovers occur, this resulted into an overall dispersive line shape~\cite{Diouf2019}. In a study on an atomic system exhibiting a preponderance of $\Lambda$-type crossovers the saturation spectrum was demonstrated to exhibit the shape of a Lamb-peak~\cite{Schmidt1994}.

\section{Results}

\begin{figure*}
\begin{center}
\includegraphics[width=0.9\linewidth]{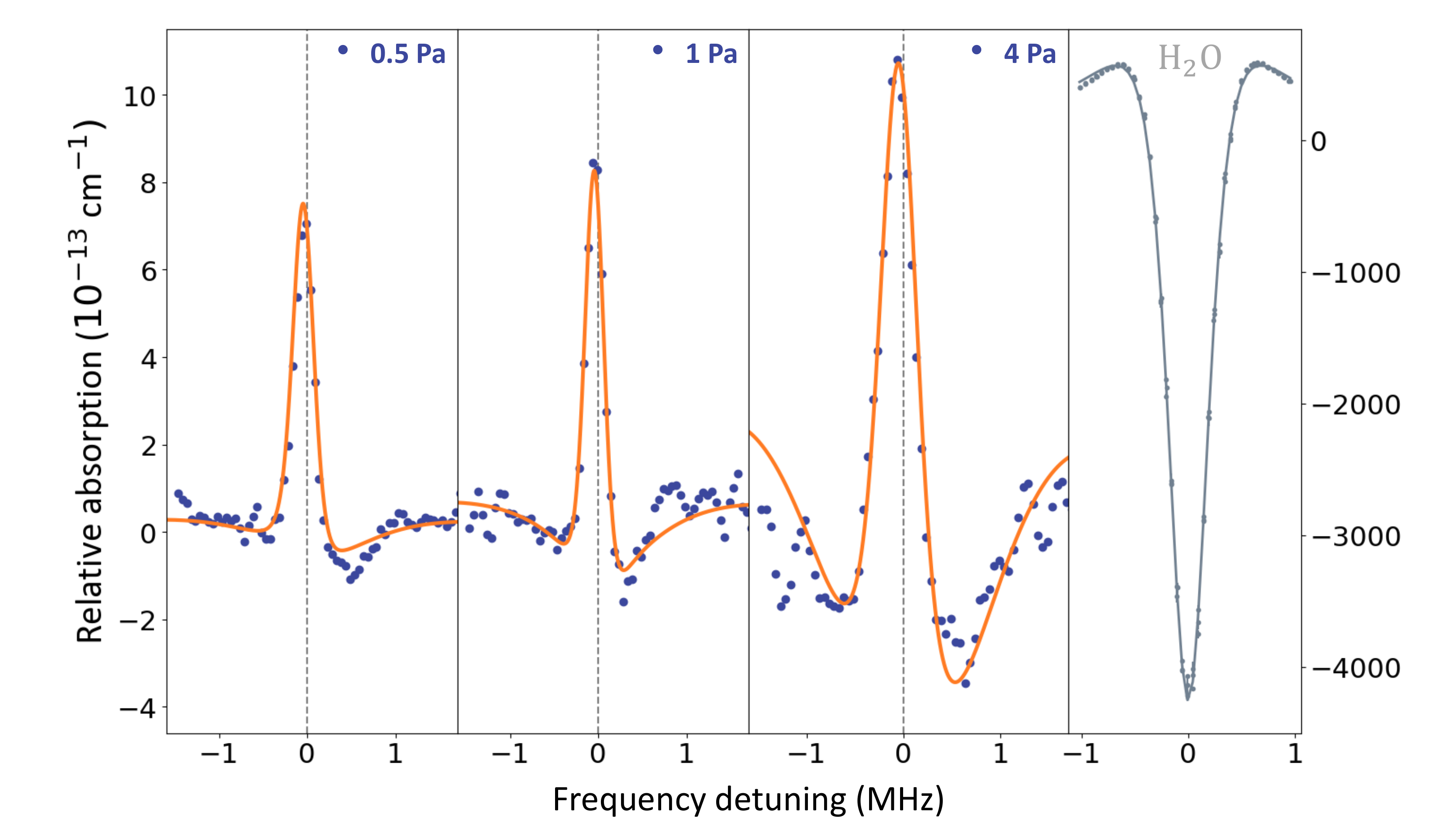}
\caption{\label{Spectra}
Recordings of saturation spectra of the P(1) ($2-0$) line in HD at different pressures as indicated. Simulated spectra from the density matrix model with optimized parameters are plotted, after convolution with a Lorentzian function to account for collisional broadening (at 45 kHz/Pa) with a full line (orange). At the right a spectrum of the water line $(000)3_{30}$ - $(101)4_{13}$ at 216,803,819,806 kHz is shown, measured under the same conditions, displaying a characteristic Lamb-dip.}
\end{center}
\end{figure*}

A few typical examples of spectra of the P(1) line of HD in the ($2-0$) band are shown in Fig.~\ref{Spectra}, where the absolute frequency is accurate to 1 kHz due to the lock of the diode laser to a frequency comb.
The observed spectra of the P(1) line display the typical characteristic of a Lamb-peak. For comparison a spectral line of H$_2$O is displayed, recorded under the same experimental conditions and settings of the electronics, thus showing a characteristic Lamb-dip and proving the opposite, Lamb-peak, nature of the HD P(1) resonance. It is noted that the NICE-OHMS technique, which is essentially a frequency-modulation spectroscopic technique, results in dispersive signals. The application of an additional slow (415 Hz in the present case) wavelength modulation of the laser is used for heterodyne detection in a lock-in amplifier, therewith improving the signal-to-noise ratio.  It produces a first derivative of this dispersive line shape~\cite{Foltynowicz2009b} as illustrated by the H$_2$O line.

\begin{figure}
\begin{center}
\includegraphics[width=0.9\linewidth]{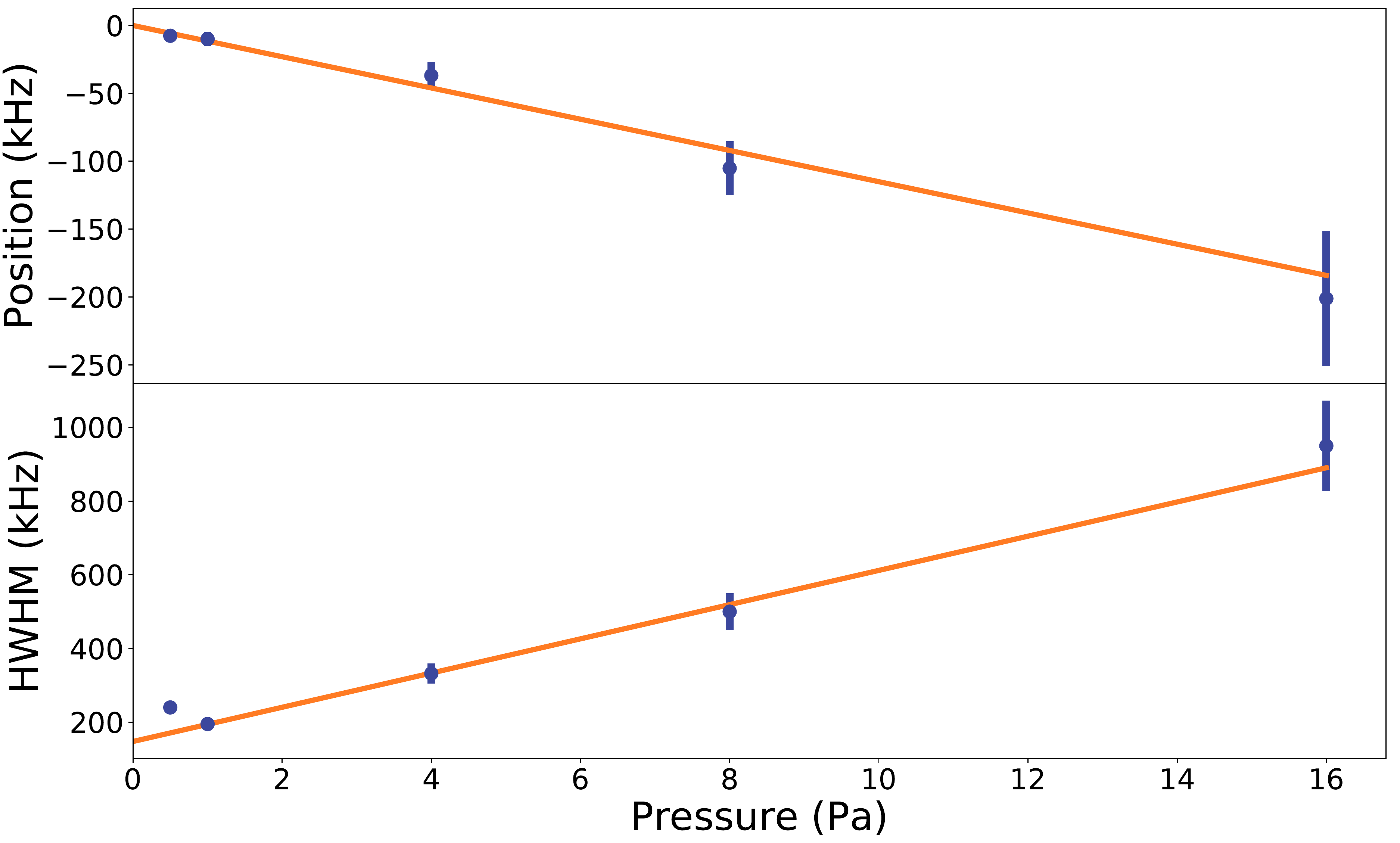}
\caption{\label{Pressure}
Transition frequencies and linewidths (HWHM) observed for the P(1) (2-0) line of HD as a function of pressure.}
\end{center}
\end{figure}

The peak positions of the lines and their widths, as obtained directly from the measurements, are plotted in Fig.~\ref{Pressure} as a function of pressure over the measurement range of [0.5 - 16] Pa.
The extracted pressure-dependent shift coefficient of $-11(1)$ kHz/Pa for P(1) is consistent with the result for R(1)~\cite{Cozijn2018}.
This shift coefficient is much larger than would be expected when using results from Doppler-broadened measurements in~\cite{Fasci2018}, and at higher pressures of $10^{4} - 10^{6}$ Pa in ~\cite{Rosasco1989} and extrapolating to the Doppler-free and Pa-level conditions in this study.
It has recently been observed that the pressure-induced line shifts in water at low pressures~\cite{Tobias2020}, using the same setup as in the present study, also do not follow the expected high-pressure trend, with the pressure shift coefficient even changing sign depending on specific transitions.
It is also worth noting that a nonlinear dependence is expected at low temperatures due to velocity-changing collisions as shown for D$_2$ in Ref.~\cite{Martinez2018}, which might be of relevance here considering the selection of cold molecules in optical saturation.

From Fig.~\ref{Pressure}, a collisional broadening parameter of 45 kHz/Pa is deduced for the saturated P(1) line. The data point for 0.5 Pa falls somewhat off the linear collision curve because that spectrum was recorded for a cavity dither peak-to-peak amplitude of $180$ kHz, while for the other lines an amplitude of $90$ kHz was used.
These results show that the width of the P(1) line is again, similar to the case of the R(1) line~\cite{Cozijn2018,Diouf2019}, narrower than expected from the contributing broadening mechanisms. A number of physical and instrumental effects contribute to the linewidth as observed, all expressed as half-width-half-maximum (HWHM) in the following. Besides the broadening caused by collisions and by amplitude modulation, the cavity-jitter, with the laser locked to the cavity, will contribute for 40 kHz, as was experimentally determined.
Intra-cavity saturation spectroscopy gives rise to recoil-doublets~\cite{Bagaev1991} separated by 68 kHz. The largest contribution to the line width is expected from the limited time for the molecules to reside in the laser beam, estimated at 400 kHz at room temperature~\cite{Borde1979}. Finally, the hyperfine structure causes broadening over the span of components, i.e. several hundred kHz.

In view of the fact that the measured linewidth is much smaller than expected from the various contributions, it is assumed that in the saturation experiment the slow molecules, that  will undergo less transit-time broadening, are preferentially detected. This is in line with the fact that the actual intracavity power (some 150 W) is much lower than the saturation parameter of the HD resonance (10 kW). Hence, in the present experiment, the saturation spectroscopy acts as a selection mechanism for cold molecules in the sample.

The line shapes as recorded for the various pressures were modeled in terms of a density matrix formalism with coupled Bloch equations as defined previously~\cite{Diouf2019},  involving populations of excited sublevels $\rho_{jj}$ and ground sublevels $\rho_{ii}$ and coherences $\rho_{ij}$:
\begin{eqnarray}
    \frac{d}{dt}\rho_{ii}= \sum_{j}\rho_{jj}\gamma_{pop,ij}-\frac{i}{\hbar}\sum_{j}(V_{ji}\rho_{ji}-\rho_{ji}V_{ij}), \\
    \frac{d}{dt}\rho_{jj}= -\sum_{i}\rho_{jj}\gamma_{pop,ij}-\frac{i}{\hbar}\sum_{i}(V_{ij}\rho_{ij}-\rho_{ij}V_{ji}), \\
    \frac{d}{dt}\rho_{ij}= -(i\Delta_{ij}+\gamma_{coh,ij})\rho_{ij}+
    \frac{i}{\hbar}V_{ij}(\rho_{ii}-\rho_{jj})\\ \nonumber
    -\frac{i}{\hbar}\sum_{k\neq j}V_{ik}\rho_{kj}, \\
    \frac{d}{dt}\rho_{jk}= (i\omega_{jk}-\gamma_{coh,jk})\rho_{jk}-\frac{i}{\hbar}(V_{ij}\rho_{ji}-\rho_{ki}V_{ki})
\end{eqnarray}
with  definitions of the Rabi frequency $V_{ij}/{\hbar}$, detuning $\Delta_{ij}=\omega_{L}-(\omega_{ij}+\vec{k}\cdot\vec{v})$, the laser frequency $\omega_{L}$, the transition frequency between  $i$ and $j$ states $\omega_{ij}$, and  the Doppler shift for the velocity class $v$ as $\vec{k}\cdot\vec{v}$.
Note that the crossover resonances, nor their estimated intensities, are explicitly included in the model; their effect is implicitly imposed when integrating over velocity space.
Decisive parameters in the model are the population relaxation rates $\gamma_{pop,ij}$ connecting $i$ and $j$ states with allowed dipole transitions and relaxation rates for coherences $\gamma_{coh,ij}$ effectively describing radiative and non-radiative processes. The refilling of the ground state from collisional decay of the excited states via $\gamma_{pop,ij}$ can be considered as a two-step process via a thermal bath.

The optical Bloch equations are solved in time steps of 1 $\mu$s for each single detuning frequency, over a span of 4 MHz approached with a step size of 1 kHz. The total integration time is fixed at 3.4 ms in order to reach the steady state solution, which must be  obtained to achieve a stable resulting spectrum. The velocity distribution of the particles is divided over 816 velocity classes which are then implemented for each single velocity in the 4000 integration steps.

In the simulations the values of $\gamma_{pop,ij}$ and $\gamma_{coh,ij}$ are considered as free parameters and are used as such to find an optimal representation of the observed line shapes.
The Rabi frequency $V/{\hbar}$ is set at $40$ kHz commensurate with the power density in the cavity. The simulation is rather insensitive to the value of $\gamma_{pop,ij}$ and it finds convergent solutions as long as it is held at $> 35$ kHz; here we have fixed $\gamma_{pop,ij}=55$ kHz.
The line shape and the effective width are sensitive to the value of $\gamma_{coh,ij}$, which is varied in the calculation to find a matching line shape.
In Fig.~\ref{Spectra} simulated spectra are plotted for an optimized parameter value for $\gamma_{coh,ij}$. For the spectrum recorded at $0.5$ Pa we find a best match for $\gamma_{coh,ij}=95$ kHz, and for the spectra at $1.0$ Pa and $4.0$ Pa the optimum is at $\gamma_{coh,ij}=125$ kHz and $\gamma_{coh,ij}=355$ kHz, respectively.
The simulations do not reproduce the line widths beyond 4 Pa ($\gamma_{coh,ij}>355$ kHz) although it retains the Lamb-peak structure.
We suspect that here another regime is reached that is dominated by collision-induced perturbations on the rovibrational levels, which are not included in the Bloch equation calculations.
After the simulation the result is convolved for the effects of recoil doublet, cavity jitter and wavelength dithering.  The resulting simulated spectra are then plotted in Fig.~\ref{Spectra} to be compared with the experimental spectra.

An important conclusion of the modeling based on this density matrix model is that an overall spectral pattern of a Lamb-peak is found, irrespective of the details and exact values of the $\gamma$-parameters invoked. For no realistic set of parameters a sign reversal could be produced. This is in agreement with the expectation (see above) that a saturation spectrum with crossover resonances only of $\Lambda$-type should produce a Lamb-peak.

From these optimized simulations, performed for each pressure, a value for the rovibrational (or hyperfine-free) transition frequency is found  as shown by the dashed line. This is done by fitting the theoretical line shape to a standard functional form (a first-order derivative of a dispersive Lorentzian~\cite{Cozijn2018}), which is subsequently fitted to the observed spectral profile. The resulting hyperfine-less transition frequency deviates somewhat from the center of the Lamb-peaks, on the order of $45$ kHz, varying by $<10$ kHz for the highest pressures. It was verified that the resulting transition frequency varies only within $10$ kHz if the parameter space of $\gamma_{coh,ij}$ is explored, where the resulting line shape and width form a selection criterion. Similarly the Rabi frequency was varied in the relevant range 35-55 kHz, imposing an additional uncertainty in the simulation of 5 kHz.
Extrapolation to zero pressure then yields the final value for the P(1) rovibrational transition. The main sources of uncertainty, related to the simulation (variation of parameter space for $\gamma_{coh,ij}$, $\gamma_{pop,ij}$ and Rabi frequency $V/{\hbar}$) and the statistics of line fitting and pressure extrapolation are combined to yield an uncertainty of $20$ kHz.
As the resulting transition frequency we find 209,784,242,007 (20) kHz.

The result of the present experimental study allows for a comparison with the theoretical result, obtained in the framework of non-adiabatic perturbation theory (NAPT)~\cite{Czachorowski2018,Komasa2019}, which is now available as a web-based on-line calculation tool~\cite{SPECTRE}. This NAPT-tool provides binding energies at accuracy of $3 \times 10^{-4}$ cm$^{-1}$, or 10 MHz, but in the calculation of transition frequencies a cancellation of uncertainties is accounted for, delivering accuracies of about $3 \times 10^{-5}$. Specifically, H2SPECTRE~\cite{SPECTRE} calculates an accuracy of 1.0 MHz for the P(1) line  and for the R-branch lines of 1.1 MHz. For the P(1) line a theoretical value of $209,784,240.1\,(1.0)$ MHz is produced~\cite{SPECTRE}. Comparison with the experimental value shows that the theoretical value is lower by $1.9$ MHz, or by $2\sigma$. Also in the previous study on the R(1) line the experimental value was higher by 1.9 MHz, again corresponding to a $2\sigma$ deviation. For the measurement of the R(2) and R(3) lines no detailed study of the hyperfine structure was performed, but the derived transition frequencies~\cite{Cozijn2018} also show an approximate $+2\sigma$ deviation from the NAPT-results. A similar deviation was found in a recent molecular beam study on HD~\cite{Fast2020}. It is noted that in all these cases the  $2\sigma$ offset is given in terms of, and is entirely due to the uncertainty in the theoretical value. This consistent deviation by $+2\sigma$ is indicative of a general and systematic offset of the NAPT-calculations for the rovibrational transitions in the (2-0) band of HD.

\section{Conclusion}

In conclusion we have demonstrated that the line shape as observed for the P(1) line in the ($2-0$) overtone band of HD can be described in terms of coherences between hyperfine substates. The same density matrix model that previously was developed to explain the observed dispersive-like spectrum of the R(1) line can now describe the shape of a single Lamb-peak for the case of P(1) with a realistic set of population and coherence decay parameters. This line shape analysis provides confidence for the extraction of the hyperfine-free or rovibrational transition frequency, at 209,784,242,007 (20) kHz, which is in agreement with the formalism of non-adiabatic perturbation theory for the hydrogen molecules within $2\sigma_{\rm theory}$.

\begin{acknowledgments}
Financial support from the  European Research Council (ERC-Advanced Grant No. 670168) and from the Netherlands Organisation for Scientific Research, via the Program “The Mysterious Size of the Proton” and the "Dutch Astrochemistry Network", is gratefully acknowledged. The authors thank the authors of Ref.~\cite{Puchalski2020} for sharing results on the hyperfine structure of HD prior to publication.
\end{acknowledgments}

%

\end{document}